\documentclass[a4paper,12pt]{article}
\usepackage[T1]{fontenc}
\usepackage[utf8]{inputenc}
\usepackage{hyperref}
\usepackage{graphicx}
\usepackage{a4wide}
\usepackage{authblk}
\usepackage{float}
\usepackage{xspace}

\usepackage{listings}
\usepackage{xcolor}
\usepackage{float}

\colorlet{punct}{red!60!black}
\definecolor{background}{HTML}{EEEEEE}
\definecolor{delim}{RGB}{20,105,176}
\colorlet{numb}{magenta!60!black}
\definecolor{grisbg}{gray}{0.95}

\lstdefinelanguage{json}{
    numbers=left,
    numberstyle=\scriptsize,
    stepnumber=1,
    numbersep=8pt,
    showstringspaces=false,
    breaklines=true,
    frame=lines,
    backgroundcolor=\color{background},
    literate=
     *{0}{{{\color{numb}0}}}{1}
      {1}{{{\color{numb}1}}}{1}
      {2}{{{\color{numb}2}}}{1}
      {3}{{{\color{numb}3}}}{1}
      {4}{{{\color{numb}4}}}{1}
      {5}{{{\color{numb}5}}}{1}
      {6}{{{\color{numb}6}}}{1}
      {7}{{{\color{numb}7}}}{1}
      {8}{{{\color{numb}8}}}{1}
      {9}{{{\color{numb}9}}}{1}
      {:}{{{\color{punct}{:}}}}{1}
      {,}{{{\color{punct}{,}}}}{1}
      {\{}{{{\color{delim}{\{}}}}{1}
      {\}}{{{\color{delim}{\}}}}}{1}
      {[}{{{\color{delim}{[}}}}{1}
      {]}{{{\color{delim}{]}}}}{1},
}

\newcommand{\includeSourceFile}[4]
{
	\lstset{basicstyle=\footnotesize\ttfamily,
			language=#4,
			float=h,
			breaklines=true,
			frame=tb,
			showspaces=false,
			showtabs=false,
			showstringspaces=false,
			tabsize=2,
			keywordstyle=\color[rgb]{0,0,1},
        	commentstyle=\color[rgb]{0.133,0.545,0.133},
        	stringstyle=\color[rgb]{0.627,0.126,0.941}
      }
	\lstinputlisting[caption=#2,captionpos=b, label=#3]{#1}
}

\newcommand{\harmony}{\textsc{Harmony}\xspace}

\begin{document}

\title{{\LARGE \bfseries The Harmony Platform}}

\author{Jean-Rémy Falleri}
\author{Cédric Teyton}
\author{Matthieu Foucault}
\author{Marc Palyart}
\author{Floréal Morandat}
\author{Xavier Blanc}
\affil{Univ. Bordeaux, LaBRI, UMR 5800, F-33400 Talence, France\\\{falleri,cteyton,mfoucaul,mpalyart,fmoranda,xblanc\}@labri.fr}

\date{}

\maketitle

\section{Context and objectives}

According to Wikipedia, 

\begin{quote}
The \emph{Mining Software Repositories} (MSR) field analyzes the rich data available in software repositories, such as version control repositories, mailing list archives, bug tracking systems, issue tracking systems, etc. to uncover interesting and actionable information about software systems, projects and software engineering.
\end{quote}

The MSR field has received a great deal of attention and has now its own research conference : \url{http://www.msrconf.org/}. However performing MSR studies is still a technical challenge. Indeed, data sources (such as version control system or bug tracking systems) are highly heterogeneous. Moreover performing a study on a lot of data sources is very expensive in terms of execution time. Surprisingly, there are not so many tools able to help researchers in their MSR quests \cite{bevan2005:Kenyon,ducasse2005:hismo,gall2009:evolizer,codeMine2013}. This is why we created the Harmony platform, as a mean to assist researchers in performing MSR studies.

\section{Overview of the Harmony platform}

The \harmony platform (\url{http://harmony.googlecode.com}) has been created to be the Swiss army knife for conducting MSR studies. Whatever your study is, we hope that \harmony will allow you to set it up quicker than you expected. For this purpose, we designed \harmony as an highly extensible platform.

Previously, we explained that most of the MSR studies have two main challenges:
\begin{itemize}
	\item They have to work with a broad set of data sources,
	\item They perform heavy computation
\end{itemize}

To cope with these issues, \harmony includes the following features:
\begin{itemize}
	\item A simple data model that abstracts the different types of data sources
	\item A set of sources extractors that can build the abstract model of a broad range of  data sources (Git, Mercurial, SVN, CVS, TFS \ldots)
	\item A collection of analyses that can be launch on the extracted data models (Object-oriented Metrics,basic statistics, \ldots).
\end{itemize}

Of course, each of these three features is extensible, meaning that you can:
\begin{itemize}
	\item Customize the data model provided by Harmony
	\item Add new data source extractors
	\item Develop your own analyses on top of the Harmony model
\end{itemize}

The cherry on top of the cake is that \harmony will take care of most of the annoying things, such as dealing with data persistence or exploiting multicore architectures.

\section{A unified model}

\begin{figure}[h]
	\centering
	\includegraphics[width=1\columnwidth]{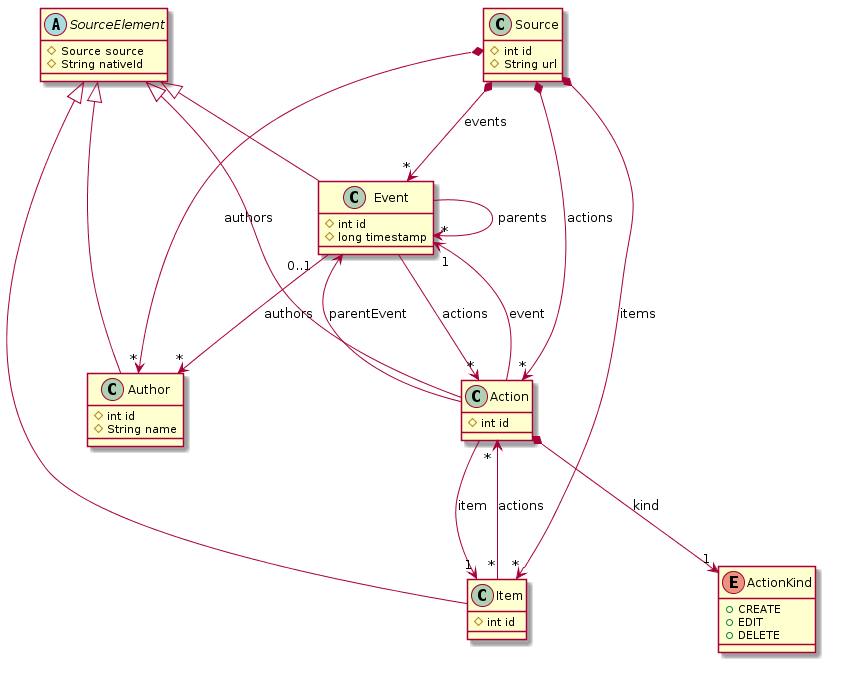}
	\caption{Data model of \harmony}
	\label{fig:data-model}
\end{figure}

\harmony provides an unified model that enables you to describe your analysis independently of any VCS. This model is "version" oriented as software evolution is a key dimension in the MSR field. The Figure \ref{fig:data-model} presents this model.

The \texttt{Source} class represents a repository. An \texttt{Event} corresponds to a specific revision of the repository. It can have multiple parent events, the \harmony model is therefore compatible with centralized or distributed versioning systems. Events are made by multiple authors : the \texttt{Author} class. Events contain a set of actions (\texttt{Action} class and the \texttt{ActionKind} enumeration) that can be considered as modifications. Each of these actions are affecting one item (\texttt{Item} class), or more precisely a file. We will not go into further details here but be aware that it is possible to extend this general model to fit the need of a specific study. The persistence of all the custom classes will also be handled by the platform, using standard JPA annotations.

Even tough this model is mainly used to abstract source repositories, it was also designed to be compatible with bug-tracking system. That is why the name of some concepts are sometimes vague. For example with a bug-tracking system, an item would be a bug.

\section{An extensible platform}
\begin{figure}[h]
	\centering
	\includegraphics[width=1\columnwidth]{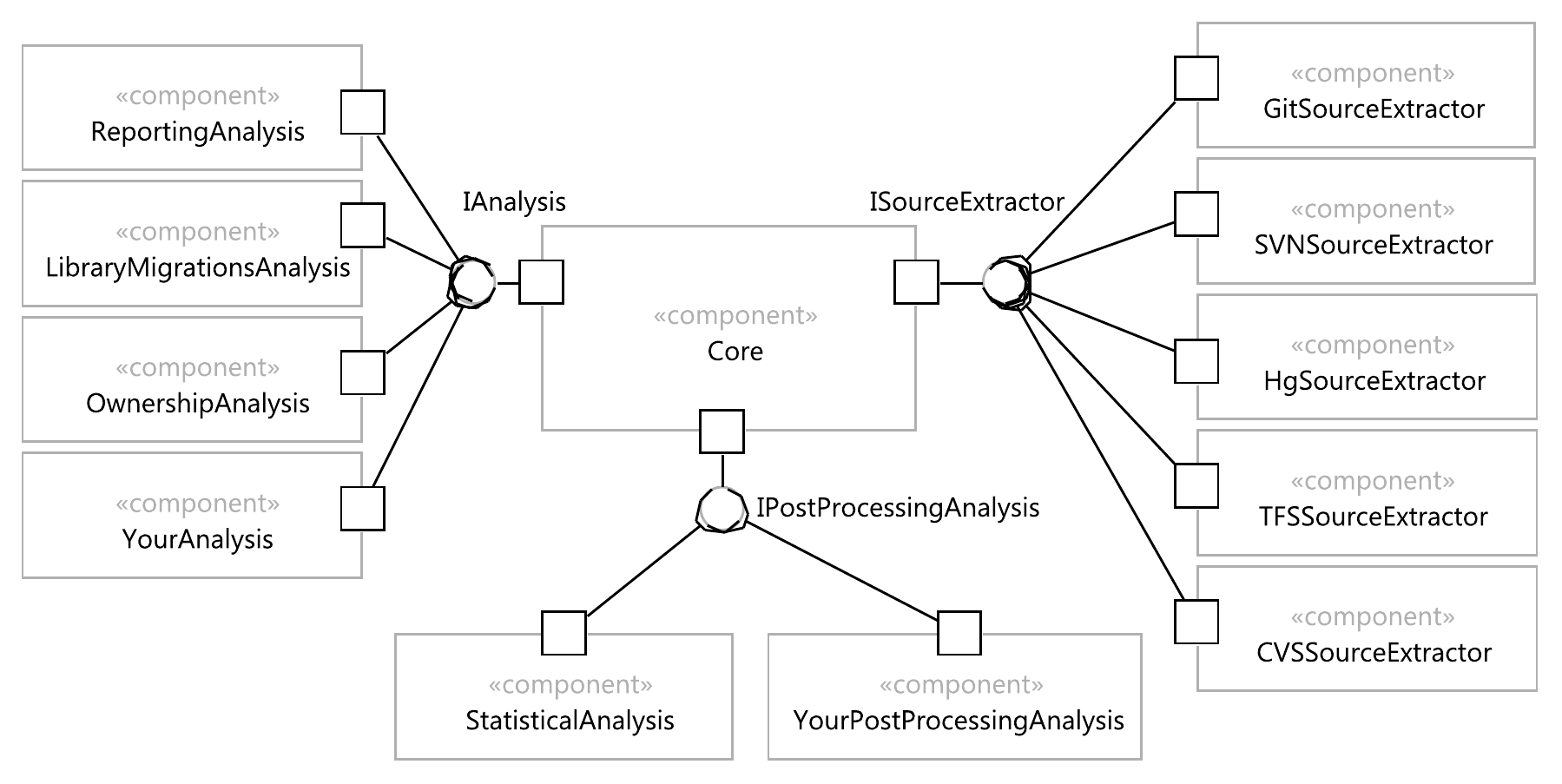}
	\caption{Architecture of \harmony}
	\label{fig:archi}
\end{figure}

The software architecture of \harmony is based on the OSGi specifications \cite{osgi} that defines a dynamic component system for the Java language. The Figure \ref{fig:archi} details this software architecture. 

At the center of the platform is the \emph{core} component that contains the definition of the abstract model, provides the standard features and defines the interfaces of the different services. Among the features provided by the \emph{core} components we find a scheduler which is in charge of executing the analyses in a correct order as well as managing parallelism. The \emph{core} component also handles data serialization to easily save your data model or exchange data between analyses. Finally the \emph{core} component embeds a collection of useful services for dealing with configuration files, output or logging. 

The \emph{core} component defines the interfaces of three services:
\begin{itemize}
	\item IAnalysis: an analysis that takes a source as input. This is the standard way for implementing an analysis. Classes that implement IAnalysis can be chained by specifying the dependencies between them in a configuration file. The scheduler will take care of executing them in a correct order. Data exchanges based on the blackboard pattern \cite{hayes1985:blackboard} can be performed by different analyses.
	\item IPostProcessingAnalysis: an analysis that take the whole collection of sources as input and that will be executed at the end. There can only be one IPostProcessingAnalysis per study.
	\item ISourceExtractor: a source extractor is in charge of building the Harmony model by exploring a repository using a particular versioning system.  
\end{itemize}

Thanks to this architecture you can develop an analysis that will be executed on a source repository no matter what versioning system it uses. In addition to the abstract model, the Harmony platform can give access to the repository files in order to perform fine-grained analyses. Developers can then easily benefit from tooling embedded in the Eclipse platform for parsing source code and configuration files such as the JDT\footnote{Java Development Tools - \url{http://www.eclipse.org/jdt/}} or CDT\footnote{C/C++ Development Tooling - \url{http://www.eclipse.org/cdt/}}.

\section{A straightforward tool}

Even though \harmony can be used with any OSGi implementation we recommend the use of the Equinox implementation \cite{gruber2005:equinox} developed by the Eclipse community. That is why we also recommend to use Eclipse as IDE in order to ease the development of your analyses. In this context, we provide an automatic installation procedures as well as a wizard for creating new analyses.

\includeSourceFile{ownership.java}{Example of analysis: computation of ownership}{code:ownership}{java}

In order to show how easy it is to develop an analysis with Harmony we illustrates it with an example. In the article \cite{bird_dont_2011} Bird et al. define that an author is a major contributor of an item if he performed at least 5\% of the actions on the files. Otherwise he is a minor contributor. We will now see how to develop an analysis with Harmony that computes the degree of ownership. After installing Harmony and using the wizard for creating a new analysis (see User Manual for details) you will just have to implements the \emph{runOn} method of the analysis class file that was generated for you by the wizard. The listing \ref{code:ownership} contains the code needed to compute the degree of ownership for each developer on each file. 

\section{Perspectives}

This papers shows that the current version of the \harmony platform already enables researchers to focus on designing and running analyses to answer research questions rather than struggling with technical details to implement them. Thanks to the modular software architecture of the \harmony platform, the situation will carry on to improve with its future versions. Components using various sampling methodologies will be developed to ease the building of representative sets of sources. It will also be possible to embed script based on the R language \cite{R2006} into analyses in order to chain them directly with standard \harmony analyses.


\bibliography{references}
\bibliographystyle{abbrv}

\end{document}